\documentclass{aa}
\usepackage{graphicx}

\def\***#1{***{\scshape #1}***}

\def\ergs{erg s$^{-1}$ cm$^{-2}$}

\begin{document}

\sloppypar

   \title{INTEGRAL detection of a long powerful burst from SLX~1735-269}

   \author{S. Molkov$^{1,2}$,  M. Revnivtsev$^{1,2}$, A. Lutovinov$^{1}$,  
 R. Sunyaev$^{1,2}$ }

   \offprints{molkov@hea.iki.rssi.ru}

   \institute{Space Research Institute, Russian Academy of
              Sciences, Profsouznaya 84/32, 117997 Moscow, Russia
        \and Max-Plank-Institute fur Astrophysik,
              Karl-Schwarzschild-Str. 1, 85740 Garching bei Munchen, Germany
        }
  \date{}

  \authorrunning{Molkov et al.}
  \titlerunning{}
        
\abstract{We present results of analysis of a bursting behavior of a low
mass X-ray binary system SLX 1735-269 during INTEGRAL observations of the
Galactic Center region in 2003. There were detected 6 type I X-ray bursts in
total, with one being much longer and more powerful then others. A strong
dependence of the bursts recurrence time on the mass accretion rate is
observed, that is likely caused by the change in the burning regime. The
long burst demonstrated a photospheric radius expansion. We discuss possible
scenarios of this long burst and show that it is unlikely a carbon burning
flash and rather a burst of large pile of hydrogen and helium accelerated by
electron capture processes in a dense accumulated layer.  
 } 
\maketitle
%

\section{Introduction}

SLX 1735-269 was discovered as a persistent X-ray source in the energy range
3-30 keV in 1985 with the Spacelab~2 mission (\cite{skinner87}).  Since then
the source was seen by different instruments with the flux level of $(2-5)
\times 10^{-10}$ ergs cm$^{-2}$ s$^{-1}$ (\cite{skinner87}, Pavlinsky et
al. 1992, 1994, \cite{zand92}, \cite{greb96}). SLX~1735-269 was detected in
hard X-rays (35--75 keV) by the SIGMA telescope on board the GRANAT
observatory in 1992 (\cite{goldw96}).  A broadband spectral analysis of the
system showed that it is likely a low mass X-ray binary
(\cite{david97}). But only in 1997 its nature was finally established after
a detection of a type I X-ray burst with a Wide Field Camera on the BeppoSAX
observatory (\cite{bazzan97}, \cite{cocchi98}). The detection of this type I
burst (unstable nuclear burning on the neutron star surface, see e.g. Hansen
\& van Horn 1975) demonstrated that SLX 1735-269 is a neutron star binary
system.

The burst detected with BeppoSAX/WFC remained the only one known burst from
this system until the launch of the INTEGRAL satellite. A large exposure
time spent by INTEGRAL on observations of the Galactic Center region allowed
us to detect more bursts from SLX 1735-269.  One of 6 detected bursts had a
duration more than 1000 sec, that is not typical for standard
hydrogen/helium type I bursts but rather similar to so-called superbursts --
unstable burning of carbon (see e.g.  \cite{cumming01}, \cite{zand04}).

In this paper we present the analysis of the bursting behavior of SLX
1735-269 concentrating on the properties of the unusual long burst.

\section{Instruments and Observations}

In this work we present results from the JEM-X monitor, module 2 (see
\cite{lund03}) and the upper layer of the IBIS telescope (ISGRI/IBIS;
\cite{uber03}) of the INTEGRAL observatory (\cite{winkl03}). These
instruments together cover a broad energy range 3--400 keV with typical
sensitivities $\sim$4 and $\sim$1.5 mCrab for one orbit ($\sim3$ days) of
observations for JEM-X (one unit) and IBIS/ISGRI, respectively.

In 2003 the Galactic Center region was observed many times with INTEGRAL
during both Open and Core Programs of the observatory (\cite{winkler03}). In
our analysis we used Galactic Center region observations of the Open Program
performed in Aug. 23 - Sep. 24, 2003 (proposal ID 0120213) and publicly
available data of Galactic Center observations performed as a part of the
Core Program in Mar. 2 -Apr. 30, 2003. A total effective exposure of SLX
1735-269 for the JEM-X telescope was $\sim 1.8$ Msec, for the IBIS telescope
-- $\sim 3.4$ Msec. The smaller effective exposure for JEM-X is caused
mainly by its smaller field of view (all observational set consists of
large number of pointings, and the optical axis of the observatory
moves significantly from pointing to pointing).

Type I bursts have a very soft X-ray spectrum and a main part of its
luminosity emitted in the energy range 3-10 keV, therefore for the search of
X-ray bursts we used the data of the JEM-X telescope, collected from the
whole detector.

JEM-X has five different telemetry formats: Full Imaging, Restricted
Imaging, Spectral/Timing, Spectral and Timing.  Usually it operates in a
``full imaging mode'', when the data are recorded as an events list with the
time resolution of 122 $\mu$sec and 256 energy channels information.  This
mode is preferable, but JEM-X has a very limited telemetry rate and
sometimes it is forced to switch in to other modes. The count rate of its
detector mainly dominated by a cosmic sources, therefore the telemetry rate
of JEM-X strongly depends on a brightness of sources within its field of
view (FOV). If the count rate of the detector exceeds some limiting value
then on-board memory buffer reserved for JEM-X is overloaded and the
information about all registered events can not be transmitted to the
Earth. So, in this cases, in order to avoid gaps in the data transmission, a
grey filtering algorithm is applied.  Depending on the detector count rate
this algorithm rejects ``n'' from ``32'' events. Thus, the real count rate
equals to the transmitted count rate multiplied by $32/(1+G)$, where G is a
grey filter value, which can varies from ``0'' to ``31''. If this algorithm
can not keep count rate below the threshold, the restricted imaging mode is
used. It has the time resolution 0.125 sec and 8 broad energy channels
covering a whole JEM-X energy band. The grey filtering algorithm can be
applied to this mode also, that leads to the significant worsening of the
time resolution.

During our observations JEM-X was mainly in a ``full imaging mode'' but a
few times it switched into a ``restricted imaging mode''

\section{Data reduction}

The analysis of the JEM-X telescope data was done using a standard INTEGRAL
Off-line Science Analysis software version 4.0 (OSA-4.0) distributed by the
INTEGRAL Science Data Center.

The JEM-X analysis package from OSA-4.0 has twelve steps of the data treatment
(see \cite{jemum}, Part II).  Below we discuss only three main levels: COR,
IMA and SPE. COR is the first level of a scientific analysis where events
after a primary telemetry pre-processing are corrected for instrumental
effects, such as an energy gain correction and positional gain correction.
An analysis up to IMA and SPE levels is based on the output data of the COR
level. After the steps IMA and SPE we have reconstructed images of the sky
and spectra of detected sources. For the following study we combined outputs
of all three levels listed above.

\begin{figure}[t]
 \vspace{-0.7cm}
 \resizebox{90mm}{!}{\hspace{-0.8cm}\includegraphics{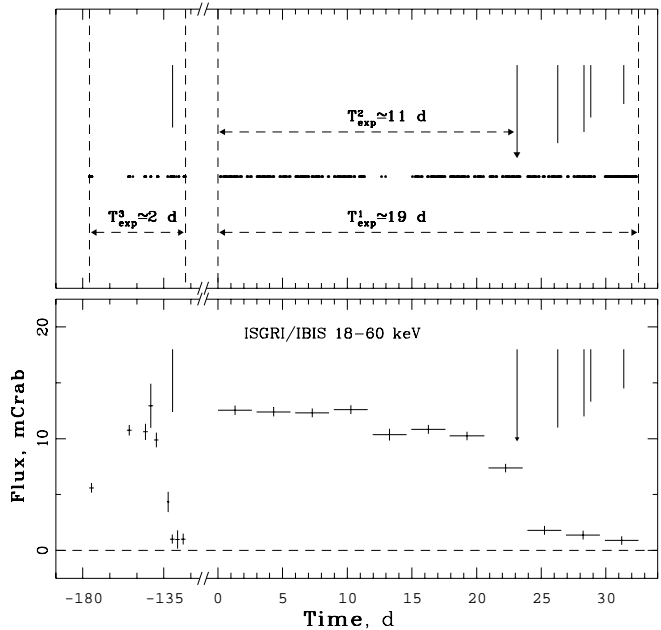}}
 \vspace{-5mm}
 \caption{History of INTEGRAL observations of SLX 1736-269 in 2003. Zero on
   the time axis corresponds to Aug. 23, 2003 -- the begining of the
   Galactic Center ultra deep observations (proposal ID 0120213) {\bf Upper
   panel:} Thick horizontal line shows time intervals when the source
   SLX~1735-269 was in FOV of the JEM-X telescope. Values above horizontal
   dashed lines with arrows show effective exposure times during different
   sets of observations. {\bf Bottom panel:} Light curve of the source
   extracted from data of IBIS/ISGRI in the 18-60 keV energy band. Each
   point represents measurement averaged over $\sim 240$ ksec (one
   revolution). Vertical solid lines on both panels indicate moments when
   type I bursts were detected. The length of this lines is proportional to
   the maximum value of the JEM-X count rate measured during the burst. The
   vertical arrow denotes the moment of the long burst beginning.}
 \label {exp}
\end{figure}

It is important to note that at this moment the OSA software is still under
the development and our investigations showed that a quantitative study of
sources is not possible from the imaging. The measured absolute values of
sources fluxes are not fully reliable while their detection and localization
can be done with a high quality. Therefore we used the JEM-X imaging
analysis only for the bursts localization.

For the timing analysis we used the data corrected for the instrumental
effects (COR level, see above). Lightcurves were constructed from a total
count rate of the JEM-X detector in required energy bands after the
correction for the grey filter factor. As the detectors background only
slightly variates with the time we used calibration observations of ``empty
fields'' in order to determine the background level. To avoid problems with
a vignetting correction in the following analysis we used only data when the
source was in the JEM-X fully coded field of view (FCFOV). Applying this
simplest procedure to the Crab observations we found that the background
subtracted count rate of Crab is stable within $\sim 20\%$. So using values
of count rates of the Crab nebula in different energy bands and the
background count rates in the same energy bands we can get the absolute flux
of a source in the FCFOV of JEM-X with a rather good accuracy.

\begin{figure}[t]
 \vspace{-1.7cm}
 \resizebox{90mm}{!}{\hspace{-0.8cm}\includegraphics{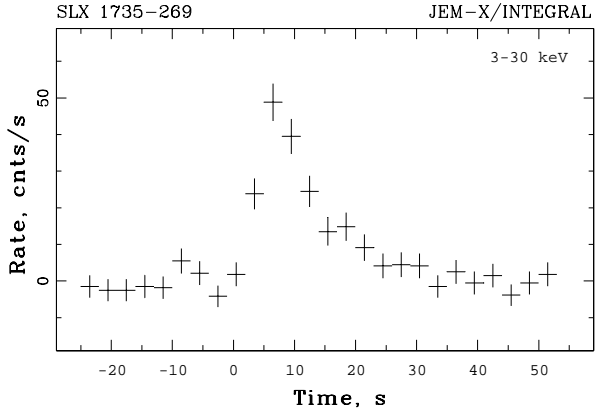}}
 \vspace{-22mm}
 \caption{Profile of the typical X-ray burst observed by INTEGRAL/JEM-X from
   SLX~1735-269 on Sept. 20, 2003 22:01:37 (UT).}
 \label {brstprof}
\end{figure}

For the spectral analysis we used an outcome of OSA-4.0 (level SPE). The
extensive study of Crab nebula observations showed that the software allows
to reconstruct the shape of the source spectrum rather well while absolute
fluxes are not reliable. Thus in the subsequent spectral analysis we used
spectra produced by OSA-4.0, but renormalize them to the fluxes obtained
from lightcurves (see above). The spectral analysis of the data obtained in
the ''restricted imaging mode'' is not possible with OSA-4.0, therefore in
this case we constructed source spectra from lightcurves in several broad
energy bands.

The IBIS/ISGRI data analysis was done with the software developed by Eugene
Churazov in Space Research Institute, Moscow (\cite{revnivtsev04}). This
software provides the absolute values of source fluxes in a wide energy band
with 10\% systematic uncertainty and allows to reconstruct the spectrum
shape with the accuracy of 3-5\%.

Data of Rossi X-ray Timing Explorer (RXTE, \cite{rxte}) observations that we
used for the construction of SLX 1735-269 broadband spectra were reduced
with the help of standard tasks of LHEASOFT 5.3 package.

\section{Results}

\begin{table}[b]
  \centering
  \caption{Bursts detected with INTEGRAL/JEM-X from the neutron star binary
    SLX 1735-269 during Galactic Center observations in 2003 
  \label{burstlist}}
  \begin{tabular}{l|c|r}
\hline
\#   &Start time, UT&Burst type\\
\hline
    1 & Apr. 15, 2003 09:48:35 & ordinary\\
    2 & Sept. 15, 2003 17:35:14 & long \\
    3 & Sept. 18, 2003 21:50:53 & ordinary\\
    4 & Sept. 20, 2003 22:01:37 & ordinary\\
    5 & Sept. 21, 2003 10:37:15 & ordinary\\
    6 & Sept. 23, 2003 23:13:05 & ordinary\\
\hline
  \end{tabular}
\end{table}

During all used observations 6 type I X-ray bursts were detected from SLX
1735-269 (see Table \ref{burstlist} and Fig.\ref{exp}). All of them with one
exception have properties similar to that observed by BeppoSAX/WFC
(\cite{bazzan97}, \cite{cocchi98}).  A typical profile of one of them (burst
number \#4 from the Table 1) is presented in Fig.\ref{brstprof}. All bursts
demonstrated an exponential decay with e-folding time of $\tau=8\pm1$
sec. Typical energies released in these bursts were approximately $E_{\rm
burst}\sim {\rm few}\times10^{39}$ ergs s$^{-1}$ (here and later we will
assume the source distance 8.5 kpc). Unfortunately a detailed spectral
analysis is not possible for these bursts due to limited statistics.

\begin{figure}[t]
 \vspace{-1.7cm}
 \resizebox{90mm}{!}{\hspace{-0.8cm}\includegraphics{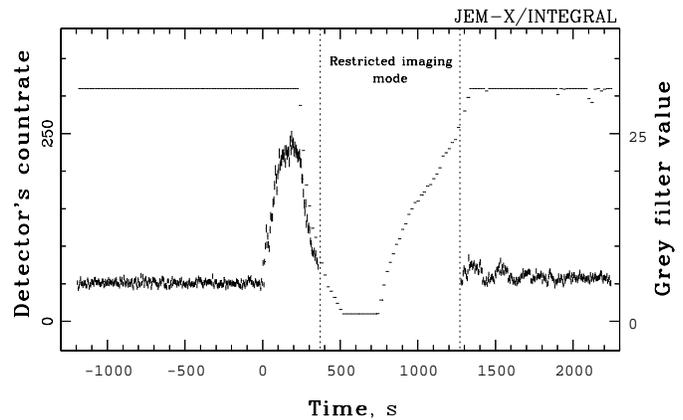}}
 \vspace{-15mm}
 \caption{Count rate collected from the whole JEM-X detector during the long
   burst. Solid step line denotes grey filter value changes. The zero time
   corresponds to the start of the burst.}
 \label {bstlcrv}
\end{figure}

An unusually long burst was observed on Sept. 15, 2003 at 17:35:14 UT during
ultra deep Galactic Center observations (pointing number 011200540010). The
count rate measured from the whole JEM-X detector and the grey filter factor
during this observation are presented in Fig.\ref{bstlcrv}. 

\subsection {Profile of the long burst}

\begin{figure*}[t]
\includegraphics[width=\textwidth]{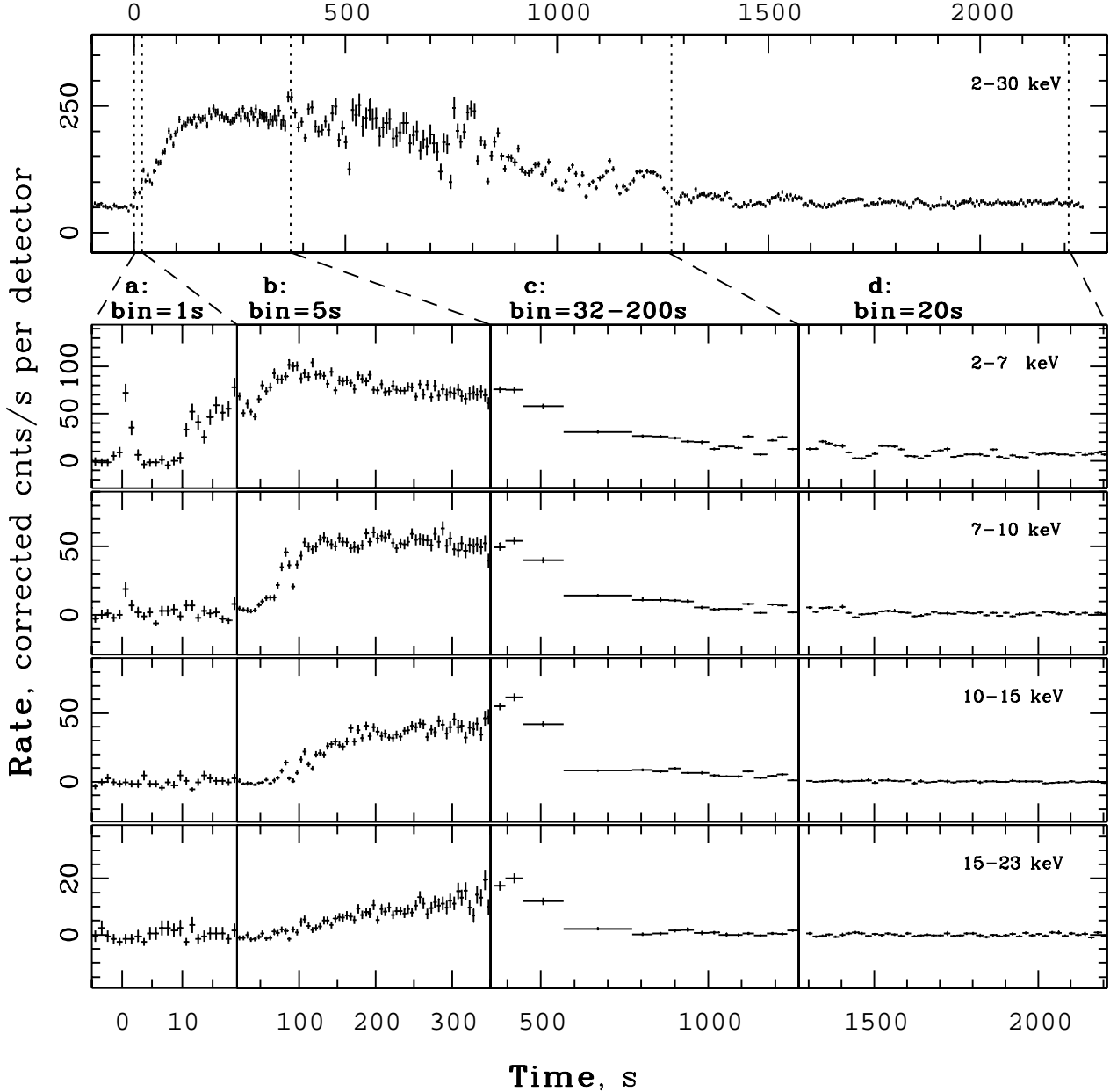}
\caption{Temporal profiles of the long burst measured by JEM-X in different
energy bands. All light curves are corrected on the grey filter factor.
Time binning for the light curve in the broad energy band (upper panel) is 5
sec for the ``full imaging mode'' and 8 sec for the ``restricted mode''.
Time binning for other curves is pointed in the figure.}
\end{figure*}

In Fig.4 we present the profile of the long burst in different energy bands
obtained from JEM-X data after the correction for the grey filter. Light
curves in the subbands are background subtracted.  The moment ``0''
corresponds to Sept. 15, 2003 17:35:14 (UT). The burst has a total duration
more than 2~ksec and its profile demonstrates several notable features. To
emphasize their we split the burst into four time intervals and plot each
interval with its own time scale. The figure shows that the burst was
started with a short burst-like event (``precursor'') that has a duration
about $\sim 2$ sec and which was more powerful in soft energy bands. During
$\sim 8$ sec after the precursor the source flux was below the detectable
level. Such ``gaps'' are typical for bursts with the photospheric radius
expansion and can be interpreted in terms of the cooling of the neutron star
photosphere during its expanding. After this ``gap'' the flux of the source
rise again to the maximum during $\sim$100-450 sec depending on the energy
band. Such behavior reflects a strong change of the source hardness and is
also typical for bursts with the photospheric radius expansion (a
contraction phase). The source intensity decay after $\sim$450 sec can be
characterized by e-folding time $\tau\sim 250$ sec in the energy band 2-7
keV and $\tau\sim150$ sec in the 15-23 keV energy band.

\begin{figure*}[t]
\includegraphics[width=\textwidth]{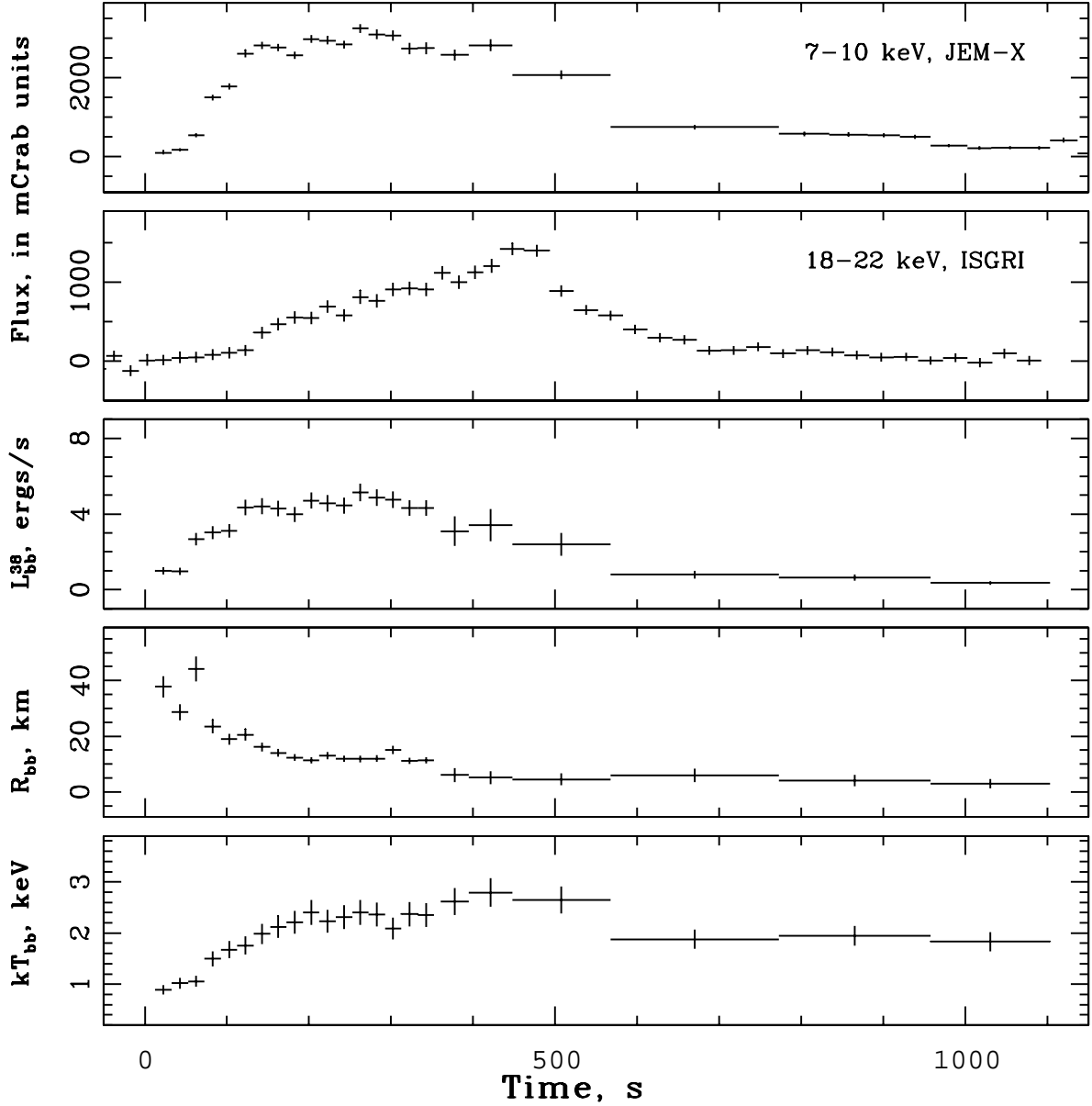}
\caption{Evolutions of SLX~1735-269 X-ray fluxes, measured by JEM-X and IBIS
  telescopes in 7-10 and 18-22 keV energy bands, respectively; its bolometric
  luminosity; the radius of photosphere and its effective temperature during
  the long burst (from the black body fit to the spectra). \label{spec_other}}
\end{figure*}

In the tail of the burst can be seen quasiperiodic oscillations of the
source flux at the time scale of $\sim 100$ sec. Such oscillations are
practically not visible in the harder energy band (10--23 keV) that
indicates that the amplitude of these variations decreases with the energy.
Interesting to note that the time scale and energy dependence of these
oscillation are very similar to those of mHz QPOs, seen in some other X-ray
bursters (\cite{revnivtsev01}). In this paper authors proposed that such
oscillations can be caused by some special regime of the nuclear burning on
the neutron star surface.

\subsection{Spectral evolution during the burst}

For the spectral analysis we extracted a source spectrum during a 2 sec time
interval covering the precursor, 17 consecutive spectra started from $\sim
20$th sec with an integration time of 20 sec, and 6 spectra in the
restricted imaging mode. All obtained spectra were fitted by a blackbody
model in the 3-20 keV energy band.

In the two upper panels of Fig.\ref{spec_other} evolutions of burst fluxes
in the JEM-X 7-10 keV and ISGRI/IBIS 18-22 keV energy bands are shown in
mCrab units. In the middle panel we present the bolometric luminosity of the
source (in units of $10^{38}$ ergs s$^{-1}$) calculated from the blackbody
spectral model as a function of time.

The fit of the spectrum obtained during the precursor by this model gives
following parameters: $T_{bb}=1.1 \pm 0.2$ keV and $R_{bb}=26 \pm 7$
km. Such large uncertainties are caused by a limited sensitivity of JEM-X
for such weak and soft events. During the next $\sim$8 sec after the
precursor the flux from the source was not detected. The evolution of the
model parameters during the following $\sim 1000$~sec is presented in two
lower panels of Fig.\ref{spec_other}.

\section{Discussion}

\subsection{Bursts and persistent flux}

Before INTEGRAL only one type I X-ray burst was detected from SLX
1735-269. This burst had a duration of $\sim$30 sec (e-fold decay time
$\sim$10 sec) and a peak flux approximately 900 mCrab (\cite{bazzan97},
\cite{cocchi98}) which translates into the energy emitted in the burst
few$\times 10^{39}$ ergs (assuming a source distance of 8.5 kpc). Such
energy release and decay time are typical for the helium burning regime (see
e.g. \cite{fujimoto81}, \cite{bildsten98}).

INTEGRAL observations of the source in 2003 contain approximately 1.8
million seconds of exposure time for the JEM-X monitor. During this exposure
we have detected in total 6 X-ray bursts. One of them was extremely powerful
and long, while others were similar to that one observed by BeppoSAX. This
long burst obviously had much longer recurrence time. No any bursting
activity was detected during $\sim$11 days (an effective exposure) before
this burst. It is important to note that there were small gaps in our
observations, but their total duration was not significant (Fig.\ref{exp}).

Several short consequent bursts allow us to estimate the burning regime
parameter $\alpha$, the ratio between an accretion energy released between
bursts to a nuclear energy released in the burst. This parameter in some way
reflects the composition of the burning fuel. Using the set of almost
consequent bursts detected from day 25 till day 33 (in the units of
Fig.\ref{exp}) we estimated $\alpha\sim100-200$ (different estimations of
$\alpha$ appear because of not clock-like bursting behavior of SLX
1735-269). Such values of $\alpha$ are typical for almost pure helium
burning (\cite{fujimoto81}, \cite{bildsten98}, \cite{woosley04}).

In order to obtain the best estimation of the source luminosity and
accretion rate we should to construct the source broadband spectrum because
the hard X-ray information alone is not a good estimator of the total
broadband or bolometric luminosity of neutron stars (e.g. \cite{barret01}).
In order to reconstruct the source broadband spectrum we included data of
the RXTE observatory in our analysis.

\begin{figure}[t]
 \vspace{-0.1cm}
 \resizebox{90mm}{!}{\hspace{-0.8cm}\includegraphics{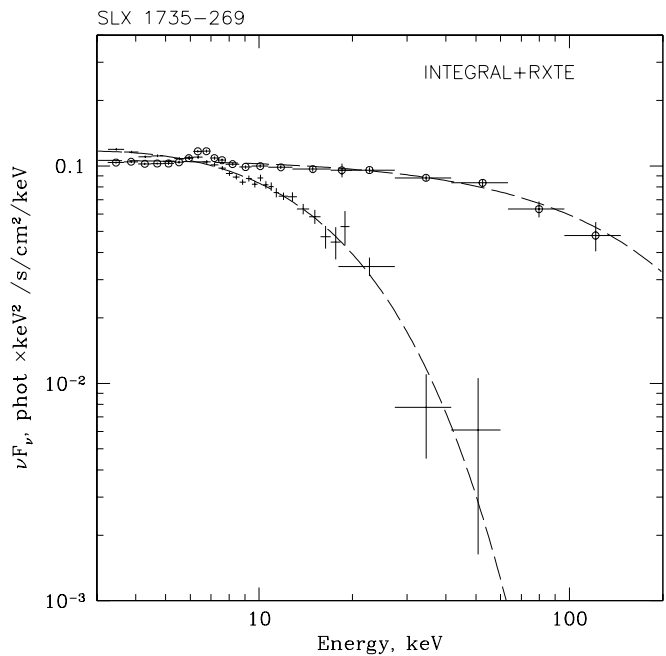}}
 \vspace{-12mm}
 \caption{Spectra of persistent X-ray emission of SLX~1735-269 at different
   states. Crosses denote the spectrum in the soft state (low hard X-ray
   flux), open circles denote the X-ray spectrum of SLX 1735-269 with high
   hard X-ray flux. All spectral points above 20 keV are based on
   INTEGRAL/IBIS data, spectral points at energies lower 20 keV taken from
   RXTE/PCA data. Dashed lines show spectra best fit models in the form of a
   power law with an exponential cutoff at high energies.}
 \label {spectra}
\end{figure}

One observation of SLX 1735-269 was performed as a part of a coordinated
INTEGRAL+RXTE observational campaign of the system. At that time the source
spectrum was hard, with a photon index $\Gamma=2.0\pm0.1$ and an exponential
cutoff at energies $E_{\rm cut}=150_{-10}^{+50}$ keV (see
Fig.\ref{spectra}).  The model broadband (0.5-100 keV) flux of the source
was $\sim 5.4\times10^{-10}$ \ergs\ that corresponds to the source
luminosity $L_{\rm x}\sim 4\times 10^{36}$ ergs s$^{-1}$. During the period
of a low hard X-ray flux the source has a completely different
spectrum. INTEGRAL observations show that the spectrum became much softer
than it was during the period of high hard X-ray flux. We searched the
public RXTE data archive in order to find the source spectrum similar to
that observed by INTEGRAL during the period of low hard X-ray (18-60 keV)
flux. RXTE observations of SLX 1735-269 performed on Feb. 14, 2003 perfectly
match this criterion. The total broadband (0.5-100 keV) flux of the source
in this state increased and the cutoff energy strongly diminished ($E_{\rm
cut}=11\pm1$ keV) in the comparison with the period of a large hard X-ray
flux (Fig.\ref{spectra}). The broadband (0.5-100 keV) flux at this state was
$\sim 6.2\times10^{-10}$ \ergs\ that corresponds to the source luminosity
$L_{\rm x}\sim 5\times 10^{36}$ ergs s$^{-1}$.

Note that the observed source luminosities differ only slightly while its
spectrum shape and burst properties differ dramatically. During the soft
state type I bursts go quite often -- the recurrence time between the bursts
is approximately 0.5-2 days. In contrast to this then the source spectrum
was very hard the recurrence time strongly increases -- $\Delta t_{\rm
b}>11$ days. This can indicate that the source persistent accretion rate is
on the border between two distinct modes of a nuclear burning. A theory says
that such transition should occur approximately at the value of $\sim 10^3$
g s$^{-1}$ cm$^{-2}$ mass accretion rate per units area
(e.g. \cite{bildsten98}).  The persistent mass accretion rate on the neutron
star in SLX 1735-269 (assuming $\eta\sim$20\% mass to energy conversion
coefficient and $L=\eta \dot{M}c^2$) is $\dot{M} \sim 2\times 10^{16}$ g
s$^{-1}$. The corresponding accretion rate per unit area is $\dot{M}\sim 1.5
\times 10^{3}$ g s$^{-1}$ cm$^{-2}$ (assuming 10 km radius of the neutron
star). Therefore indeed we have the system in which the accretion rate is on
the border of two regimes of the nuclear burning. Note that this supports
the assumption of the source distance 8.5 kpc.

Short bursts with e-folding time $\tau\sim10$ sec are rather typical among
type I bursters. Very long bursts, with a duration more than 1000 sec are
rare and a present theory provides two possible scenarios for such bursts --
the burning of a large pile of hydrogen and helium fuel, or the carbon flash
(see e.g. \cite{kuulkers02}).

\subsection{Carbon flash scenario}

The source accretion rate per unit area $\dot{M}\sim 1.5\times 10^{3}$ g
s$^{-1}$ cm$^{-2}$ does not allow to burn of He in a stable manner
(\cite{bildsten98}). Therefore the accumulation of a large amount of carbon
can not be done via this process and the most of the carbon fuel needed to
produce the burst should be accreted from a companion star. An estimation of
the energy released in the burst gives $E_{\rm burst}\sim 2\times10^{41}$
ergs. To provide such energy with the carbon fuel (a nuclear energy release
$\simeq 5.6\times10^{17}$ ergs g$^{-1}$, Cumming \& Bildsten 2001) we should
burn $M\sim 5\times10^{23}$ g of carbon. To accumulate such amount of carbon
via accretion from the main sequence secondary star we need more than $10^9$
sec (assuming a solar abundance of $^{12}C$ in the accreted
matter). Therefore the probability to detect one burst in our set of
observations (an effective exposure time $\sim 2\times10^6$ sec) is less
than one tenth of percent. Thus we conclude that the long burst from SLX
1735-269 is very unlikely a carbon one.

\subsection{He/H flash scenario}

The theory says that for such low accretion rate an X-ray burst can be
caused by a mixed hydrogen and helium burning triggered by a hydrogen
ignition accelerated by electron capture processes (see e.g.
\cite{fushiki92}, \cite{bildsten98}, \cite{bc98}, \cite{kuulkers02}). The
nuclear energy release per gramm of such fuel is $\sim {\rm few} \times
10^{18}$ ergs g$^{-1}$ (depending on the fuel composition). We should to
burn $\sim 10^{23}$ g, which will be accumulated after $\sim 3\times 10^{6}$
sec of the accretion with the above mentioned rate. This recurrence time is
comparable with the exposure of our observations. The critical surface
density necessary to make electron capture processes very effective is or
the order of $y\sim 10^{10}$ g cm$^{-2}$, that means that the accreted mass
$M\sim 10^{23}$ g of the matter should be involved. This value again is
compatible with the observed value of the mass burned in the long burst.

The length of the burst in this scenario mainly depends on the limiting
(Eddington) luminosity. It is governed by the energy budget of the burst
divided by the Eddington luminosity on the neutron star. A stability of the
source luminosity during a few hundreds of seconds at the peak of the burst
(Fig.\ref{spec_other}) supports this assumption.

Summarizing all of the above we can conclude that the long burst detected
from SLX 1735-269 most likely is a result of the unstable burning of a large
pile of mixed hydrogen and helium accelerated by electron capture
processes. It is interesting to note that the similar long burst was
observed by BeppoSAX from another faint neutron star binary SLX 1737-282
(\cite{zand02}).

\bigskip

{\it Acknowledgements.} Authors thank to E.Churazov for the developing of
the methods of the analysis of the IBIS data and software. We would like to
thank Prof. L.Bildsten for very useful discussions and comments. Research
has made use of data obtained through the INTEGRAL Science Data Center
(ISDC), Versoix, Switzerland, Russian INTEGRAL Science Data Center (RSDC),
Moscow, Russia, and High Energy Astrophysics Science Archive Research Center
Online Service, provided by the NASA/Goddard Space Flight Center. This work
was supported by RFBR grant 04-02-17276, grants of Minpromnauka
NSH-2083.2003.2 and 40.022.1.1.1103.

\end{document}